\documentclass[conference, 9pt]{IEEEtran}
\IEEEoverridecommandlockouts

\usepackage{cite}
\usepackage{amsmath,amssymb,amsfonts}
\usepackage{algorithm}
\usepackage{algorithmic}
\usepackage{graphicx}
\usepackage{textcomp}
\usepackage{xcolor}
\usepackage{subcaption}

\usepackage{colortbl}
\usepackage{dhucs} 
\usepackage{booktabs} 
\usepackage{multirow}
\usepackage{comment}
\usepackage{lineno}
\usepackage{color} 
\usepackage{setspace} 
\usepackage{graphicx} 
\usepackage{makecell}
\usepackage{enumitem} 
\usepackage{relsize}
\usepackage{tabularx}
\usepackage{diagbox}

\newcommand\korean[1]{}

\newcommand\refFigure[1]{Fig.~\ref{#1}}
\newcommand\refTable[1]{Table~\ref{#1}}

\newcommand\refAlgo[1]{Algorithm~\ref{#1}}

\newcommand\blind[1]{XXXX}

{\renewcommand{\baselinestretch}{0.92}

\def\BibTeX{{\rm B\kern-.05em{\sc i\kern-.025em b}\kern-.08em
    T\kern-.1667em\lower.7ex\hbox{E}\kern-.125emX}}
\begin{document}

\title{\LARGE\relsize{+1.5} HH-PIM: Dynamic Optimization of Power and Performance with Heterogeneous-Hybrid PIM for Edge AI Devices\\
}


\author{
Sangmin Jeon$^{\dag}$, Kangju Lee$^{\dag}$, Kyeongwon Lee, and Woojoo Lee$^{*}$ \\
School of Intelligent Semiconductor Engineering, Chung-Ang University, Seoul, Korea \\
\{jademin96, agl0312, since69se, space\}@cau.ac.kr

\thanks{
This paper has been accepted for publication at the 62nd Design Automation Conference (DAC 2025). This document represents the camera-ready version.

$^{\dag}$ Sangmin Jeon and Kangju Lee contributed equally to this work.

$^{*}$Woojoo Lee is the corresponding author.}
}

\maketitle

\begin{abstract}
Processing-in-Memory (PIM) architectures offer promising solutions for efficiently handling AI applications in energy-constrained edge environments. 
While traditional PIM designs enhance performance and energy efficiency by reducing data movement between memory and processing units, they are limited in edge devices due to continuous power demands and the storage requirements of large neural network weights in SRAM and DRAM. 
Hybrid PIM architectures, incorporating non-volatile memories like MRAM and ReRAM, mitigate these limitations but struggle with a mismatch between fixed computing resources and dynamically changing inference workloads. 
To address these challenges, this study introduces a Heterogeneous-Hybrid PIM (\textit{HH-PIM}) architecture, comprising high-performance MRAM-SRAM PIM modules and low-power MRAM-SRAM PIM modules. 
We further propose a data placement optimization algorithm that dynamically allocates data based on computational demand, maximizing energy efficiency. 
FPGA prototyping and power simulations with processors featuring HH-PIM and other PIM types demonstrate that the proposed HH-PIM achieves up to 60.43\% average energy savings over conventional PIMs while meeting application latency requirements. 
These results confirm HH-PIM’s suitability for adaptive, energy-efficient AI processing in edge devices.
\end{abstract}


\section{Introduction}
\normalsize

With the advent of artificial intelligence (AI), real-world applications are rapidly expanding, fueling a trend to embed AI capabilities into IoT devices across diverse fields. 
However, traditional server-centric data processing, such as cloud computing, faces significant energy and latency challenges due to processing and communication overloads. 
Consequently, distributing AI workloads to edge devices has become a promising solution, with recent research focusing on enabling on-device AI through TinyAI models that support lightweight, local computations~\cite{shafique2021tinyml,dong2023model}.


In energy-constrained edge environments, integrating Processing-in-Memory (\textit{PIM}) architectures has emerged as a promising approach for executing AI applications efficiently~\cite{zheng2023accelerating, yang2023memory, chih202116,heo2022t, Park:TCAS24, lee2024radar}. 
PIM enhances performance and energy efficiency in memory-intensive tasks, such as AI applications, by minimizing the overhead of data movement between processing and memory units. 
Early PIM designs primarily employed volatile memories like SRAM~\cite{su202015} and DRAM~\cite{lee2021hardware}, but these designs faced challenges in storing large neural network weights due to the continuous power demands of volatile memory and issues such as SRAM leakage power and the periodic refresh cycles required by DRAM. 
To address these limitations, PIM architectures based on non-volatile memories (\textit{NVMs}), such as MRAM~\cite{chiu202322nm, cai202333} and ReRAM~\cite{xue2021cmos, yang2021pimgcn}, were proposed. 
These designs achieve high energy efficiency in weight storage, especially when combined with power-gating techniques. 
However, NVMs may introduce additional read/write latency, potentially impacting overall neural network performance. 
Consequently, recent advances in PIM design have introduced hybrid architectures combining SRAM and NVM, known as Hybrid-PIM (\textit{H-PIM}). 
These architectures use NVM to store weight data and SRAM as a buffer for input and output data, thereby enhancing both performance and energy efficiency~\cite{zhu2023pim,kim2023recpim,zhang2024efficient}.


However, H-PIM faces distinct limitations in achieving optimal energy efficiency during dynamic scenarios where inference loads fluctuate in real time on edge devices.
For instance, an edge device running a YOLO model for real-time object detection experiences substantial variations in processing demand depending on the number of objects detected per video frame. 
Operating at a fixed performance level across all time intervals—typically set for peak computational load—inevitably leads to inefficient energy consumption. 
To address this, we observe that the mismatch between fixed computing resources and dynamically changing workloads has long been a challenge in traditional CPU-centric architectures. 
Established solutions, such as Dynamic Voltage and Frequency Scaling (\textit{DVFS})~\cite{Lee:DATE14, panda2022energy, Lee:TCAD15, bouzidi2023hadas} and heterogeneous multi-processor architectures with high-performance and low-power cores~\cite{Lee:COOCHIPS16,Park:DATE23,Choi:JSA24}, have been extensively researched for this purpose. Our focus centers on heterogeneous architectures, as DVFS continues to face significant challenges in edge devices due to added design complexities, such as DC-DC converters and real-time power monitoring. 
In contrast, heterogeneous architectures, exemplified by ARM’s big.LITTLE architecture~\cite{biglittle}, are widely implemented in commercial processors, effectively improving energy efficiency by adapting to dynamic computational loads.
Building on this insight, we propose configuring PIM modules, which integrate memory and Processing Elements (\textsl{PEs}) for independent computation, into high-performance and low-power configurations. 
This approach allows PIM architectures to dynamically balance performance and energy consumption throughout the application runtime.


To further elaborate this idea, we introduce Heterogeneous-Hybrid PIM (\textit{HH-PIM}), an architecture designed to dynamically optimize performance and energy efficiency for AI applications on edge devices. 
HH-PIM integrates two distinct PIM modules: a High-Performance (\textit{HP}) PIM module and a Low-Power (\textit{LP}) PIM module. 
Each module’s memory consists of a hybrid configuration of MRAM and SRAM banks, resulting in four types of memory: \textsl{HP-MRAM}, \textsl{HP-SRAM}, \textsl{LP-MRAM}, and \textsl{LP-SRAM}. 
Unlike conventional hybrid PIM architectures—where NVM is primarily allocated for weight storage and SRAM is reserved as an input-output buffer—HH-PIM adopts an adaptive approach. During periods of high computational demand, HH-PIM actively utilizes SRAM for weight storage as well, maximizing responsiveness to fluctuating inference loads. 
Additionally, to capitalize on HH-PIM’s architecture, we propose an optimal data distribution algorithm that minimizes energy consumption by dynamically adjusting data allocation across the four memory types. 
This combinatorial optimization algorithm allocates weight data across HP-MRAM, HP-SRAM, LP-MRAM, and LP-SRAM to reduce energy use while balancing workload between HP-PIM and LP-PIM. 
With this design, HH-PIM efficiently adapts to the changing computational demands of AI applications, achieving significant energy savings without compromising performance.


To verify the functionality and evaluate the effectiveness of our proposed technique, we modeled the memory and PEs and performed an RTL-level design of the entire PIM processor, incorporating the HH-PIM architecture. 
We then conducted FPGA prototyping to confirm correct operation and measure performance of the developed PIM processor, while power measurements were obtained through synthesis using 45nm process technology and memory model simulations. 
Through various benchmark scenarios, we experimentally validated the energy-saving potential of the proposed HH-PIM architecture and data distribution algorithm. 
Results showed that our approach maximizes energy efficiency while satisfying the latency requirements of AI applications in edge computing environments. Specifically, the developed PIM processor demonstrated superior adaptability to real-time inference load variations compared to conventional PIM-based processors, achieving up to 60.43\% average energy savings across different benchmark scenarios relative to the baseline processor. 
These outcomes validate the suitability of the HH-PIM architecture for maximizing efficiency in edge processors running AI applications.


\section{HH-PIM Architecture for AI Edge Processors}

\begin{figure}
\centerline{\includegraphics[width=0.48\textwidth]{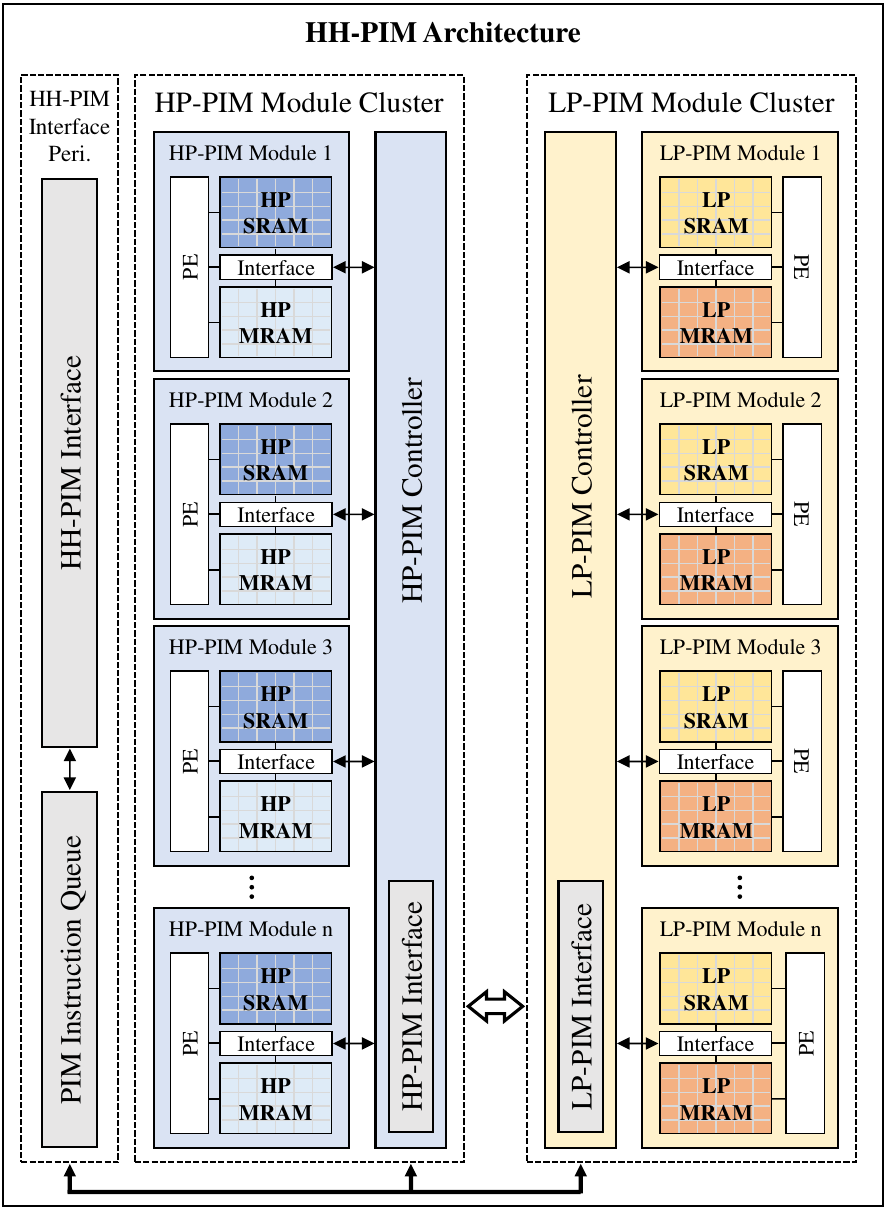}}
\caption{Block diagram of the proposed HH-PIM architecture.}
\label{fig:hhpim}
\end{figure}

\refFigure{fig:hhpim} illustrates the proposed HH-PIM architecture, designed as a solution to dynamically maximize energy efficiency in response to changes in inference workload, while meeting the performance demands of AI applications on edge devices.
Basically, HH-PIM adopts the general structure of near-memory computing architectures, as suggested in previous research~\cite{lee2021hardware, kim2022aquabolt}. 
It comprises multiple PIM modules, each with PEs and memory banks, a controller that manages these modules, and an interface for external communication. Operating based on dedicated PIM instructions, commands received from the processor core are sequentially stored in the \textit{PIM Instruction Queue}. 
Unlike conventional PIM architectures, where a single controller manages all PIM modules based on PIM instructions, HH-PIM incorporates two distinct controllers: \textsl{HP-PIM Controller} and \textsl{LP-PIM Controller}, as shown in the figure. 
This dual-controller setup is designed specifically to support HH-PIM's unique heterogeneous architecture, which consists of two types of clusters: HP-PIM module cluster operating at high performance with higher power consumption, and LP-PIM module cluster operating at low power with reduced performance. 
The HP-PIM and LP-PIM Controllers are responsible for controlling and synchronizing their respective module types.


Another notable feature of HH-PIM is its hybrid memory architecture, where the PIM module’s memory consists of MRAM and SRAM with distinct characteristics and read/write latencies. 
The previous H-PIM architectures leverage the data characteristics of neural networks by storing large weight data in NVM to enhance energy efficiency, while using SRAM or DRAM for input/output data to improve computational performance by providing low-latency access.
HH-PIM builds upon this approach, exploiting the performance and energy efficiency benefits of hybrid memory. 
However, to overcome the limitations of fixed memory allocation based solely on data characteristics, HH-PIM adds flexibility. When computational demand spikes and maximum PIM computing power is required, HH-PIM allows weight data to be stored in SRAM, thus enabling stable operation even in cases where traditional H-PIM architectures cannot meet the maximum latency requirements of certain applications.
To realize this capability, the PIM module is designed to support variable operand counts retrieved from MRAM and SRAM during computation. 
Each PIM module’s internal interface is controlled by the controller, which dynamically adjusts the load process based on data storage status. 
This design ensures synchronization of differing memory read cycles and access speeds between MRAM and SRAM in the LOAD state, guaranteeing reliable operation.


This heterogeneous and hybrid architecture of HH-PIM enables flexible adaptation to real-time variations in inference workloads for AI applications. 
When computational demand is high, the HP-PIM module and SRAM can be actively utilized to boost processor performance. 
Conversely, in low-demand scenarios, computations can be maximally allocated to the LP-PIM module to meet application latency requirements and maximize energy efficiency. 
To implement this strategy effectively, it is essential to design the PIM Controllers with precision, as they play a crucial role by synchronizing components that operate at different speeds and minimizing data movement overhead between PIM operations.

\begin{figure}
\hskip 3pt
\centerline{\includegraphics[width=0.48\textwidth]{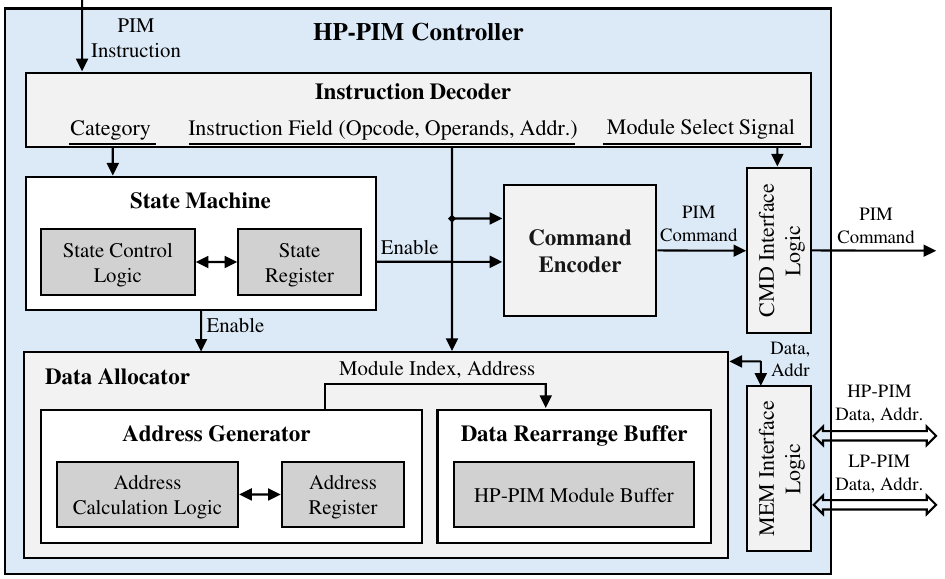}}
\caption{Block diagram of the HP-PIM Controller architecture.}
\label{fig:con}
\vskip -4pt
\end{figure}

The architecture of HP-PIM and LP-PIM Controllers is fundamentally identical.
\refFigure{fig:con} shows the architecture block diagram of the HP-PIM Controller, which consists of a \textsl{State Machine}, \textsl{Instruction Decoder}, \textsl{Command Encoder}, \textsl{Data Allocator}, and \textsl{Interface Logic}. 
The controller operates through the basic PIM instruction cycle, which includes the FETCH-DECODE-LOAD-EXECUTE-STORE phases, managed internally by the State Machine. 
The Instruction Decoder decodes the fetched instruction into components such as the instruction type (\textit{Category}), specific operation or data movement details to be executed by the PIM module (\textit{Instruction Field}), and the target module for the operation (\textit{Module Select Signal}). 
The Command Encoder then generates command signals for each PIM module based on the decoded instruction details.


To minimize overhead from frequent data movement between HP-PIM and LP-PIM modules, as well as between MRAM and SRAM within each PIM module—resulting from the dynamic energy optimization of the HH-PIM architecture—the controller design of HH-PIM incorporates a Data Allocator and separates the Interface Logic as seen in \refFigure{fig:con}. 
The Data Allocator manages data placement to minimize external data movement during PIM operations for various computations, such as convolution and Multiply-Accumulate (MAC) operations within AI applications. 
Additionally, the Interface Logic is divided into a CMD Interface Logic for delivering commands to the PIM modules and a MEM Interface Logic for data movement between PIM modules. 
When operand data is appropriately positioned in each PIM module’s memory, computations can proceed efficiently through the CMD Interface Logic alone.
Furthermore, the bandwidth of the MEM Interface Logic is scaled according to the number of PIM modules within each cluster, enabling parallel data transfers across clusters from multiple PIM modules simultaneously.


The controller’s operation sequence for data movement between PIM modules is as follows. 
First, when a data placement instruction from the core prompts data transfer between HP-PIM and LP-PIM modules, the controller decodes the instruction and provides the relevant operation and address information to the Data Allocator. 
Based on the provided addresses, the Data Allocator accesses the memory in each PIM module via the MEM Interface Logic and stores the data to be transferred in the \textsl{Data Rearrange Buffer}. 
The Data Rearrange Buffer retains the data until the destination PIM module in the opposite cluster is ready for data writing, preventing data conflicts caused by the speed discrepancy between HP-PIM and LP-PIM modules. 
Next, the \textsl{Address Generator} within the Data Allocator generates the destination PIM module index and memory bank addresses in the opposite cluster to facilitate data movement. 
The generated addresses are then provided to the Data Rearrange Buffer, enabling it to place the buffered data into the memory banks of the designated PIM modules.


In summary, the PIM controllers in HH-PIM are meticulously designed to support data transfers between HP-PIM and LP-PIM modules and to effectively manage operations within each PIM module. This design maintains operational consistency and prevents data conflicts, supporting the HH-PIM architecture in achieving its design goals of real-time performance and energy efficiency optimization.


\section{Dynamic Data Placement Strategy for Optimizing Energy Efficiency in HH-PIM}

Optimal data placement in the HH-PIM architecture is essential for achieving maximum energy efficiency while ensuring smooth application execution without latency. 
In HH-PIM, each layer of a neural network is distributed across HP-PIM and LP-PIM modules for parallel computation, with the final output obtained by aggregating results from each module. 
When high performance is prioritized, it is critical to minimize idle times where the HP-PIM module completes its tasks quickly but must wait for the slower LP-PIM module. 
To achieve low output latency, a balanced distribution of data between HP-PIM and LP-PIM modules, as well as an optimal utilization ratio of MRAM and SRAM—each with different read/write latencies—must be carefully managed.

On the other hand, energy efficiency can be prioritized when AI applications do not always require the lowest possible inference latency. 
In such cases, HH-PIM can operate below peak performance by allocating more data to the energy-efficient LP-PIM module and increasing the usage of MRAM, thus improving overall energy efficiency while maintaining acceptable application latency. 
Based on this approach, we propose a dynamic data placement strategy that periodically adjusts the distribution of weight data during runtime. This strategy optimizes energy efficiency by shifting computational loads according to real-time demands, focusing on maximizing energy savings while meeting the minimum latency requirements of the application.


\subsection{Problem Definition}
In the proposed weight data placement strategy, data redistribution occurs every predefined interval $T$, referred to as the \textit{time slice}. 
We define a set of PIM operations generated by a single inference process in an AI application as a \textit{task}. 
These tasks, generated by inference, are stored in a task buffer, and HH-PIM sequentially processes the tasks stored from the previous \textit{time slice} within the current \textit{time slice}. 
This approach ensures that the operational latency of HH-PIM does not exceed $2T$, thereby maintaining the inference latency required by AI applications.
In this framework, the number of tasks stored in the task buffer determines the minimum time $t_{constraint}$ needed for each task to be processed within the \textit{time slice} without delay. 
Specifically, each task’s processing time $t_{task}$ must not exceed $t_{constraint}$ to guarantee that all tasks are completed within the allotted \textit{time slice}, ensuring the system adheres to the required maximum latency.

The proposed method aims to achieve the most energy-efficient data distribution for each task within the time constraint $t_{constraint}$. 
The HH-PIM architecture includes four distinct storage spaces: HP-MRAM, HP-SRAM, LP-MRAM, and LP-SRAM, each with different power consumption and latency characteristics. 
Consequently, we are faced with a discrete combinatorial optimization problem of determining the optimal placement of each weight data in order to minimize energy consumption.
This problem can be reduced to a well-known \textit{knapsack problem}~\cite{kellerer2004knapsack}, with necessary adaptations for our specific context. 
Here, each storage space corresponds to an item $i$ in the knapsack problem, and the time constraint $t_{constraint}$, task execution time $t_{task}$, and energy consumption per task $E_{task}$ represent the knapsack’s capacity, item weight, and item value, respectively. 
However, this problem is more complex than the standard knapsack problem, as it exhibits the characteristics of both an unbounded knapsack (due to the multiple selections of a particular storage space) and a multi-choice knapsack (due to the fixed total number of selected storage spaces, equivalent to the number of weight data).


The mathematical formulation of this problem is as follows:
\begin{align}\label{eqn:problem}
 \tag{1}
& \text{Minimize}~E_{task} = \sum\limits_{i=1}^n e_i \cdot x_i \\\notag
& \text{subject to:}~\sum\limits_{i=1}^n t_i \cdot x_i \leq t_{constraint},~~\sum\limits_{i=1}^n x_i = k, \\ \notag
& ~~~~~~~~~~~~~~x_i \in \mathbb{Z}_{\geq 0},~~\forall i = 1, 2, \cdots , n. \notag
\end{align}
where $n$ denotes the number of storage spaces, $x_i$ represents the number of data assigned to storage space $i$, and $k$ is the total number of weight data to be stored. 
Additionally, $t_i$ and $e_i$ denote the computation time per weight and the energy consumption per weight in storage space $i$, respectively. 
This fomula defines the objective function to minimize energy consumption per task, \( E_{task} \), along with constraints on execution time and the total count of selected items.

\subsection{Proposed Solution for Optimized Data Placement}
Since this problem is inherently more complex than the standard knapsack problem, it is naturally NP-hard. 
To address the optimization problem defined in (\ref{eqn:problem}), we propose a bottom-up dynamic programming (DP) approach. 
This method uses a DP table to store intermediate results, avoiding redundant calculations and solving the problem incrementally. 
The recurrence relation is defined as follows:
\begin{equation}\label{eqn:recur}
\tag{2}
\begin{split}
dp[i][t][k] = 
\begin{cases} 
dp[i-1][t][k], & (t_i \cdot k > t) \\[5pt]
\min\left( dp[i-1][t][k], \right. \\
\left. dp[i][t - t_i][k-1] + e_i \right), & (t_i \cdot k \leq t)
\end{cases}
\end{split}
\end{equation}
where $i$, $t$, and $k$ denote the current storage space being considered, the time constraint, and the number of weights, respectively. 
The value $dp[i][t][k]$ represents the minimum energy consumption required to store exactly $k$ weight data across the first $i$ storage spaces while satisfying the time constraint $t$.
In this equation, when $t_i \cdot k > t$, storing $k$ weights in the current storage space exceeds the time constraint. 
Therefore, in this case, the value from the previous storage space $dp[i-1][t][k]$ is carried forward. 
Conversely, if $t_i \cdot k \leq t$, the recurrence considers the minimum between $dp[i-1][t][k]$ and $dp[i][t - t_i][k-1] + e_i$, where the latter term represents the energy consumption if one additional weight is stored in the current storage space $i$, based on the previous results.

Meanwhile, in implementing the solution based on (\ref{eqn:recur}), it is essential to consider the architectural constraints of HH-PIM. 
HH-PIM allows for parallel execution of weight data distributed between HP-PIM and LP-PIM clusters; however, parallel processing is not feasible for weights stored across MRAM and SRAM within each module. 
To address this, we propose a partitioned approach where the DP algorithm is applied separately to HP-PIM and LP-PIM clusters, generating individual DP tables for each cluster. 
By combining these tables, we identify the optimal allocation of weights that minimizes energy consumption within the time constraint, yielding the best combination of $k$ values, ($k_{\textit{hp}}$, $k_{\textit{lp}}$), where $k_{\textit{hp}}$ and $k_{\textit{lp}}$ represent the number of weights assigned to HP-PIM and LP-PIM clusters, respectively. 
This approach effectively accounts for both inter-cluster parallelism and intra-module serialization, optimizing energy efficiency while adhering to the operational constraints of HH-PIM.


\begin{algorithm}[t]
\scriptsize\relsize{+0.5}
\caption{Finding Optimal Data Placement with DP.}\label{alg:dp}
\begin{algorithmic}[1]
\STATE \textbf{function} KNAPSACK\_MIN\_ENERGY
\STATE \quad Set \textit{dp}[\textit{i}][\textit{t}][\textit{k}] = $\infty$ and \textit{count}[\textit{i}][\textit{t}][\textit{k}] = 0 for all \textit{i}, \textit{t} and \textit{k}.
\STATE \quad Set \textit{dp}[\textit{i}][\textit{t}][0] = 0 for all \textit{i} and \textit{t}.
\STATE \quad \textbf{for} \textit{i} from 1 to $n / 2$:
\STATE \qquad \textbf{for} \textit{k} from 1 to \textit{K}:
\STATE \qquad \quad \textbf{for} \textit{t} from 1 to \textit{T}:
\STATE \qquad \qquad \textbf{if} \textit{t\(_\textit{i}\)} $\leq$ \textit{t} \textbf{then}
\STATE \qquad \qquad \quad Calculate the possible previous state time index \textit{t}$-$\textit{t\(_\textit{i}\)}
\STATE \qquad \qquad \quad \textit{dp}[\textit{i}][\textit{t}][\textit{k}] = $\min$(\textit{dp}[\textit{i}$-$1][\textit{t}][\textit{k}], \textit{dp}[\textit{i}][\textit{t}$-$\textit{t\(_\textit{i}\)}][\textit{k}$-$1]$+$\textit{e\(_\textit{i}\)})
\STATE \qquad \qquad \quad Update \textit{count}[\textit{i}][\textit{t}][\textit{k}] based on the chosen minimum path.
\STATE \qquad \qquad \textbf{else}
\STATE \qquad \qquad \quad \textit{dp}[\textit{i}][\textit{t}][\textit{k}] = \textit{dp}[\textit{i-1}][\textit{t}][\textit{k}]
\STATE \qquad \qquad \quad \textit{count}[\textit{i}][\textit{t}][\textit{k}] = \textit{count}[\textit{i-1}][\textit{t}][\textit{k}]
\STATE \qquad \qquad \textbf{end if}
\STATE \qquad \quad \textbf{end for}
\STATE \qquad \textbf{end for}
\STATE \quad \textbf{end for}
\STATE \textbf{end function}
\end{algorithmic}
\end{algorithm}
\normalsize

First, \refAlgo{alg:dp} presents the pseudo-code of the algorithm based on (\ref{eqn:recur}). 
Here, $K$ denotes the total number of weight data to be stored in HH-PIM, and \textit{count}[\textit{i}][\textit{t}][\textit{k}] is a variable used to trace the path to the optimal energy state. 
Lines 2 and 3 set the base conditions for the recurrence relation, while lines 7-13 correspond to the recurrence relation defined in (\ref{eqn:recur}). 
Notably, because \refAlgo{alg:dp} is performed separately for both HP-PIM and LP-PIM, the iteration for $i$ in line 4 only proceeds up to $n/2$.


Next, \refAlgo{alg:comb} finds the optimal combination of $k$ values that minimizes energy consumption using the results of the DP tables \textit{dp\(_\textit{hp}\)} and \textit{dp\(_\textit{lp}\)} generated for HP-PIM and LP-PIM, respectively. 
Here, \textit{min\_energy} denotes the minimum energy consumption, while \textit{k\_opt\_hp} and \textit{k\_opt\_lp} represent the number of weights assigned to HP-PIM and LP-PIM at this minimum. 
As a result, \refAlgo{alg:comb} yields \textit{allocation\_state}[\textit{n}][\textit{t}], representing the distribution state of weight data for each time constraint. 
In the proposed method, both \refAlgo{alg:dp} and \refAlgo{alg:comb} are performed only once during the application initialization phase to construct a Look-up Table (\textit{LUT}) for the final output, \textit{allocation\_state}. 
This LUT allows rapid determination of the optimal weight placement state for varying $t_{constraint}$ values required at each \textit{time slice} during application runtime. 
Furthermore, the calculation of $t_{constraint}$ at runtime incorporates the data movement overhead time needed for transitioning from the previous \textit{time slice} data allocation to the new one, ensuring no inference delay arises due to data movement overhead.


\begin{algorithm}[t]
\footnotesize
\caption{Finding Optimal ($k_{\textit{hp}}$, $k_{\textit{lp}}$).}\label{alg:comb}
\begin{algorithmic}[1]
\STATE \textbf{function} SET\_ALLOCATION\_STATE
\STATE \quad \textbf{for} \textit{t} from 0 to \textit{T}:
\STATE \quad Set \textit{min\_energy} $=$ $\infty$, \textit{k\_opt\_hp} = $0$, \textit{k\_opt\_lp} = $0$
\STATE \qquad \textbf{for} \textit{k\(_\textit{hp}\)} from 1 to \textit{K} :
\STATE \qquad \quad \textit{k\(_\textit{lp}\)} = $K$ $-$ \textit{k\(_\textit{hp}\)}
\STATE \qquad \quad \textbf{if} \textit{dp\(_\textit{hp}\)}[\textit{$n / 2$}][\textit{t}][\textit{k\(_\textit{hp}\)}] $+$ \textit{dp\(_\textit{lp}\)}[\textit{$n / 2$}][\textit{t}][\textit{k\(_\textit{lp}\)}] $<$ $\infty$ \textbf{then}
\STATE \qquad \qquad \textbf{if} \textit{min\_energy} $>$ \textit{dp\(_\textit{hp}\)}[\textit{$n / 2$}][\textit{t}][\textit{k\(_\textit{hp}\)}] $+$ \textit{dp\(_\textit{lp}\)}[\textit{$n / 2$}][\textit{t}][\textit{k\(_\textit{lp}\)}] \textbf{then}
\STATE \qquad \qquad \quad \textit{min\_energy} $=$ \textit{dp\(_\textit{hp}\)}[\textit{$n / 2$}][\textit{t}][\textit{k\(_\textit{hp}\)}] $+$ \textit{dp\(_\textit{lp}\)}[\textit{$n / 2$}][\textit{t}][\textit{k\(_\textit{lp}\)}]
\STATE \qquad \qquad \quad \textit{k\_opt\_lp} $=$ \textit{k\(_\textit{lp}\)}
\STATE \qquad \qquad \textbf{end if}
\STATE \qquad \quad \textbf{end if}
\STATE \qquad \quad \textbf{if} \textit{min\_energy} remains $\infty$, \textbf{continue} to next \textit{t}
\STATE \qquad \quad \textit{k\_opt\_hp} $=$ \textit{K} $-$ \textit{k\_opt\_lp}
\STATE \qquad \quad Set \textit{allocation\_state}[\textit{n}][\textit{t}] with current \textit{k\_opt\_hp} and \textit{k\_opt\_lp}
\STATE \qquad \textbf{end for}
\STATE \quad \textbf{end for}
\STATE \textbf{end function}
\end{algorithmic}
\end{algorithm}
\normalsize

The time complexities of \refAlgo{alg:dp} and \refAlgo{alg:comb} are $O(n \cdot T \cdot K)$ and $O(T \cdot K)$, respectively. 
Although these algorithms are executed only once during application initialization, the full \textit{time slice} range for $T$ and the large number of weight data $K$, often tens to hundreds of thousands in lightweight neural network models, may impose considerable overhead for energy optimization across all points in an edge processor. 
To mitigate this, we limit the resolution of optimization to ensure that the total computation time does not exceed 1\% of each \textit{time slice}, thereby avoiding excessive fine-grained calculations. 
The resulting optimal data distribution and improvements in energy efficiency are presented in the following section, along with detailed experimental settings.


\section{Experimental Work}\label{sec:exp}
\subsection{Development and Evaluation of PIM Processors for Power and Performance Assessment}

\begin{table}[t]
\centering
\vskip 5pt
\caption{Developed specifications for HH-PIM and other PIM architectures.}
\resizebox{\columnwidth}{!}{
\begin{tabular}{ccc}
        \toprule
        \textbf{Architecture} & \textbf{PIM Module Configuration} & \textbf{Memory Types (per module)}  \\ \midrule
        \textbf{Baseline-PIM} & 8 HP-PIM & 128kB SRAM  \\ 
        \textbf{Heterogeneous-PIM} & 4 HP-PIM + 4 LP-PIM   & 128kB SRAM\\ 
        \textbf{Hybrid-PIM} & 8 HP-PIM & 64kB MRAM + 64kB SRAM\\ 
      \textbf{HH-PIM} & 4 HP-PIM + 4 LP-PIM & 64kB MRAM + 64kB SRAM  \\ \bottomrule
    \end{tabular}
}
\label{tab:comparison}
\vskip 2pt
\end{table}

To evaluate the effectiveness of the proposed HH-PIM architecture and its data placement optimization strategy, we first implemented the HH-PIM architecture from \refFigure{fig:hhpim} at the RTL level. 
The HP and LP clusters were each configured with four HP-PIM and four LP-PIM modules, with each PIM module equipped with 64kB of MRAM and SRAM. For a fair and precise comparison, we established Baseline-PIM, Heterogeneous-PIM (Hetero.-PIM), and Hybrid-PIM as comparison groups, each also implemented at the RTL level. 
\refTable{tab:comparison} presents the specifications of each comparison group.


\begin{figure}[t]
\vskip -8pt
\centerline{\includegraphics[width=0.48\textwidth]{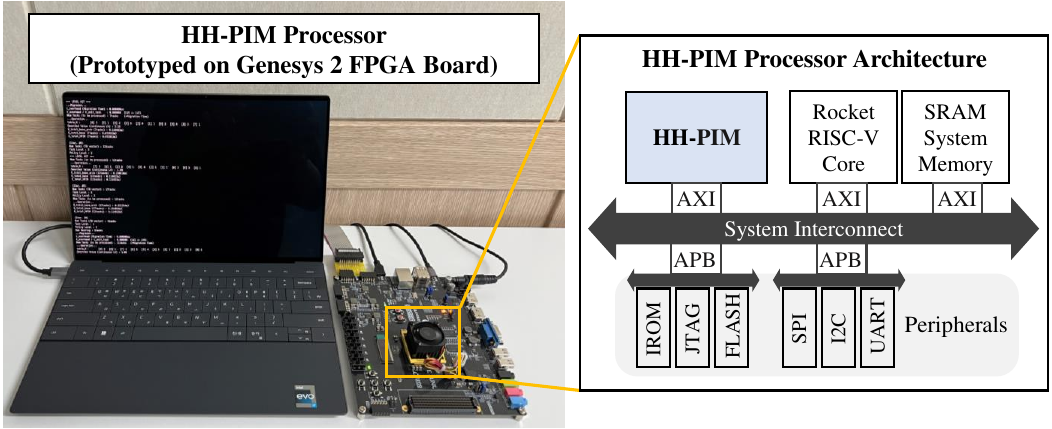}}
\caption{Design and FPGA prototyping of the HH-PIM processor.}
\label{fig:soc}
\vskip -4pt
\end{figure}

\begin{table}[t]
\setlength{\tabcolsep}{10pt} 
\caption{FPGA prototype resource utilization.}
\vskip -2pt
\centering
\resizebox{0.9\columnwidth}{!}{
\begin{tabular}{c|c|c|c|c}
\toprule
\textbf{IPs}                        	& \textbf{LUTs}	& \textbf{FFs} 	& \textbf{BRAMs} 	& \textbf{DSPs}  	\\ \midrule
RISC-V Rocket Core               	& 14,998           	& 9,762            	& 12                     	& 4                      	\\ \midrule
Peripherals                             	& 4,704             	& 7,159            	& -                       	& -                      	\\ \midrule
System Interconnect               	& 5,237             	& 7,720            	& -                      	& -                       	\\ \midrule
HP-PIM Module                      	& 968                & 1,055           	& 32                 		& 2                        	\\
HP-PIM Module Controller       	& 2,823             & 875              	& -                  		& -                       	\\ \cmidrule(l){2-5}
Total (HP-PIM module cluster) 	& 6,951        	& 5,460    		& 128 		     	& 8             		\\ \midrule
LP-PIM Module                      	& 1,074        	& 1,094    		& 32		      		& 2     		    	\\
LP-PIM Module Controller        	& 2,149    		& 875    		& -      			& -    	            	\\ \cmidrule(l){2-5}
Total (LP-PIM module cluster)  	& 6,680 	        & 5,616   		& 128	      		& 8	                 	\\ \bottomrule
\end{tabular}
}
\vskip -2pt
\label{tab:fpga_resources}
\end{table}

\begin{table}
\centering
\caption{Latency comparison of HP-PIM and LP-PIM modules.}
\vskip -2pt
\resizebox{0.79\columnwidth}{!}{
\begin{tabular}{c|cc|cc|c}
\toprule
\multirow{2}{*}{\textbf{ \diagbox{}{~~Latency\\ ~~($ns$)}}} & \multicolumn{2}{c|}{\textbf{STT-MRAM}} & \multicolumn{2}{c|}{\textbf{SRAM}} & \multirow{2}{*}{\textbf{PE}} \\
\cmidrule{2-5}
 & \textbf{Read} & \textbf{Write} & \textbf{Read} & \textbf{Write} &  \\
\midrule
\textbf{HP-PIM ($V_{dd}=1.2V$)} & 2.62 & 11.81 & 1.12 & 1.12 & 5.52\\
\midrule
\textbf{LP-PIM ($V_{dd}=0.8V$)} & 2.96 & 14.65 & 1.41 & 1.41 & 10.68\\         
\bottomrule
\end{tabular}
}
\label{tab:latency}
\vskip -6pt
\end{table}

Subsequently, we developed processors equipped with the proposed HH-PIM and each of the comparative PIM architectures. 
RTL design and simulation were conducted using RISC-V eXpress (RVX)~\cite{han2020developing,Jang:ACCESS21}. 
The diagram on the right side of \refFigure{fig:soc} illustrates the architecture of the processor equipped with HH-PIM, which is built with a single RISC-V Rocket~\cite{Rocket} core. 
To facilitate efficient transmission and reception of large-scale AI application data, the HH-PIM communicates with the core through the AXI protocol, offering high bandwidth and low latency. 
For the system interconnect, we utilized ${\mu}NoC$~\cite{Han:ISLPED19}, a lightweight Network-on-Chip optimized for edge devices.

\begin{table}[t]
\vskip -8pt
\caption{TinyML model specs and PIM operation ratios.}
\vskip -2pt
\centering
\resizebox{0.95\columnwidth}{!}{
\begin{tabular}{ccccc}
\toprule
\textbf{Model}         & \textbf{\# Param} & \textbf{\# MAC}     & \textbf{PIM Operation} & --  \\ \midrule
EfficientNet-B0        & 95k           & 3.245M    & 85\%           & \multirow{3}{*}{\parbox{2.5cm}{\centering INT8 Quantized\\ \& Pruned}}\\
MobileNetV2            & 101k          & 2.528M      & 80\%           &                \\
ResNet-18              & 256k          & 29.580M   & 75\%           &               \\ \bottomrule
\end{tabular}
}
\label{tab:tinyml_models}
\end{table}

\begin{figure}[t]
\vskip 2pt
    \centering
    \begin{subfigure}{0.48\columnwidth}
        \centering
        \includegraphics[width=\linewidth]{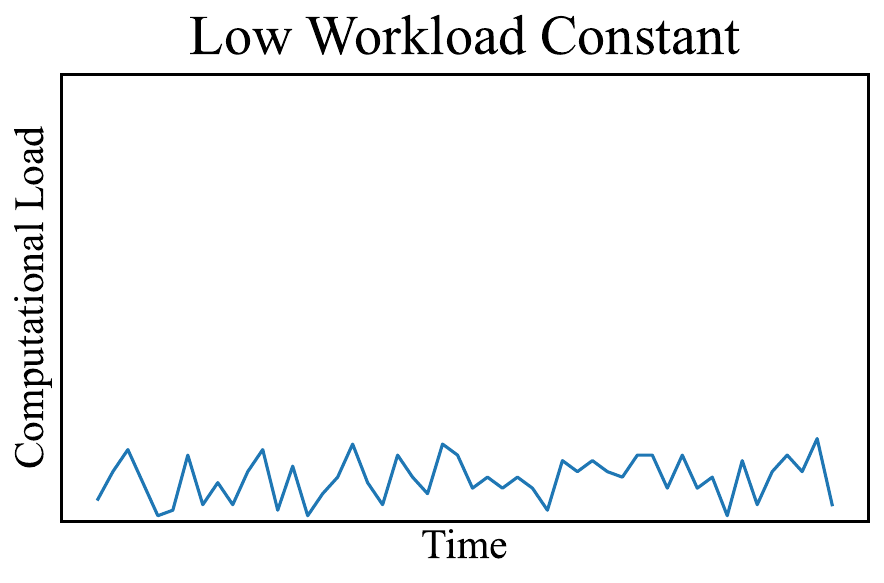}
        \vspace{-7mm}
        \caption{Case 1}
        \label{fig:low-workload-constant}
    \end{subfigure}
    \hfill
    \begin{subfigure}{0.48\columnwidth}
        \centering
        \includegraphics[width=\linewidth]{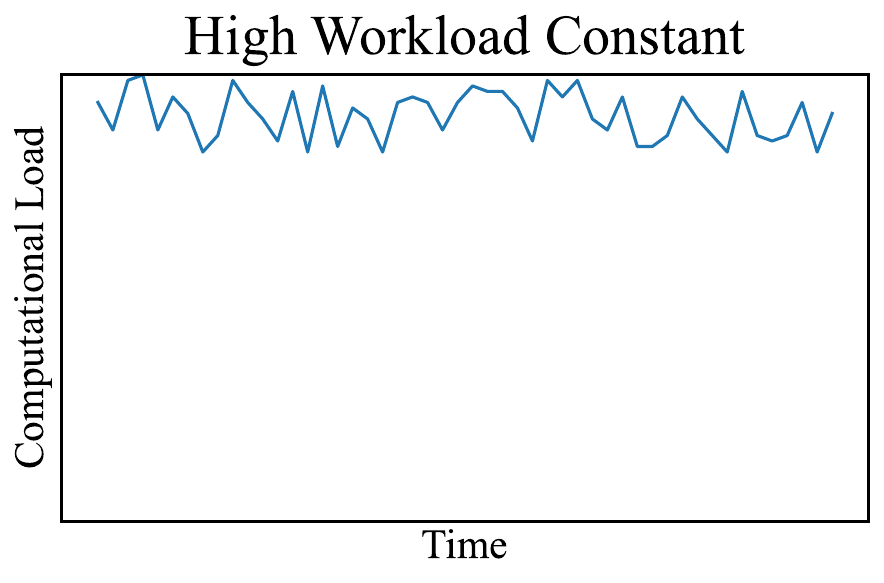}
        \vspace{-7mm}
        \caption{Case 2}
        \label{fig:high-workload-constant}
    \end{subfigure}
    
    \begin{subfigure}{0.48\columnwidth}
        \centering
        \includegraphics[width=\linewidth]{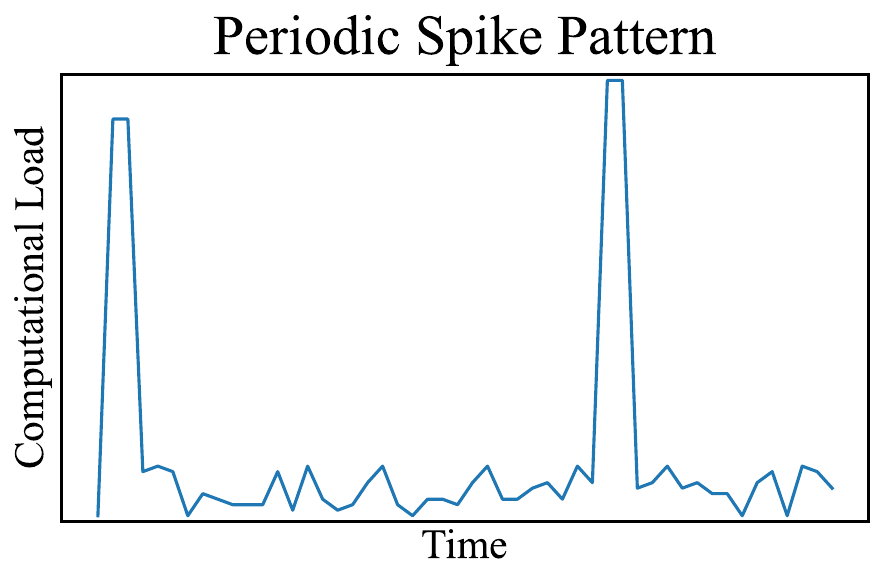}
        \vspace{-7mm}
        \caption{Case 3}
        \label{fig:periodic-spike}
    \end{subfigure}
    \hfill
    \begin{subfigure}{0.48\columnwidth}
        \centering
        \includegraphics[width=\linewidth]{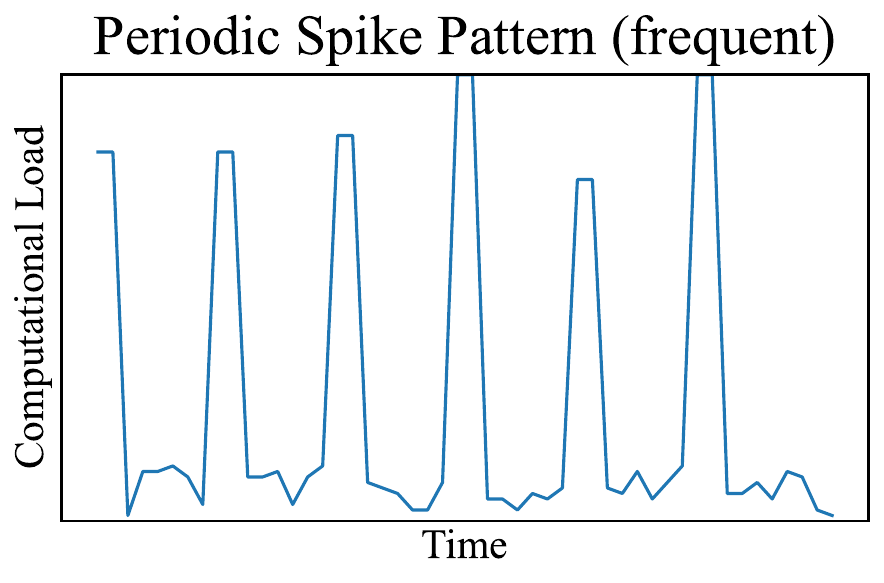}
        \vspace{-7mm}
        \caption{Case 4}
        \label{fig:periodic-spike-frequent}
    \end{subfigure}
    
    \begin{subfigure}{0.48\columnwidth}
        \centering
        \includegraphics[width=\linewidth]{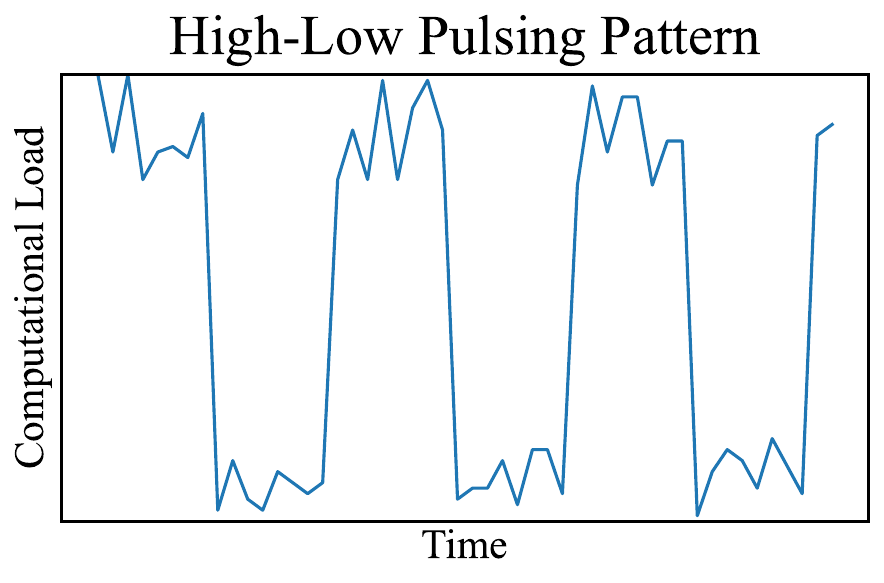}
        \vspace{-7mm}
        \caption{Case 5}
        \label{fig:high-low-pulsing}
    \end{subfigure}
    \hfill
    \begin{subfigure}{0.48\columnwidth}
        \centering
        \includegraphics[width=\linewidth]{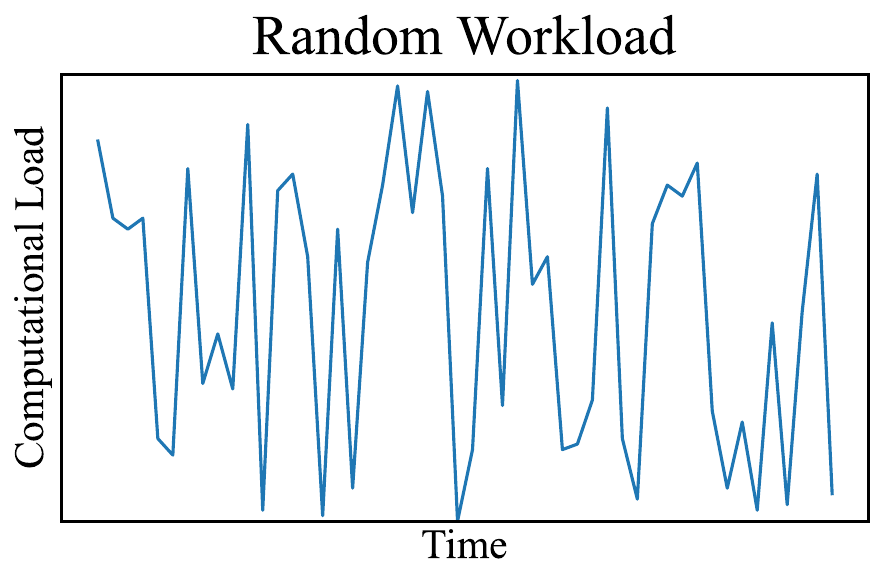}
        \vspace{-7mm}
        \caption{Case 6}
        \label{fig:random-workload}
    \end{subfigure}

    \caption{Various workload scenarios of the AI benchmark app.}
    \label{fig:workload-patterns}
    \vskip 2pt
\end{figure}

\begin{table}[h!]
\centering
\vskip 2pt
\caption{Power consumption ($mW$) across memory types in HP-PIM (1.2V) and LP-PIM (0.8V).}
\vskip -2pt
\resizebox{\columnwidth}{!}{
\begin{tabular}{c|cc|cc|cc}
\toprule
\multirow{3}{*}{\makecell{\diagbox{}{\textbf{Power} \\ \textbf{}}}} & \multicolumn{2}{c|}{\textbf{STT-MRAM}} & \multicolumn{2}{c|}{\textbf{SRAM}} & \multicolumn{2}{c}{\textbf{PE}} \\
\cmidrule{2-7}
 & \makecell{\textbf{Dynamic} \\ \textbf{(Read/Write)}} & \textbf{Static}  & \makecell{ \textbf{Dynamic} \\ \textbf{(Read/Write)}}& \textbf{Static} & \textbf{Dynamic} & \textbf{Static} \\
\midrule
 \makecell{ \textbf{HP-PIM}}& 428.48 / 133.78 & 2.98 & 508.93 / 500 & 23.29 & 0.9 & 0.48 \\
\midrule
 \makecell{ \textbf{LP-PIM}}& 179.05 / 47.78 & 0.84 & 177.3 / 177.3 & 5.45 & 0.51 & 0.25 \\            
\bottomrule
\end{tabular}
}
\label{tab:power}
\end{table}

Next, we performed functional verification and performance evaluation through FPGA prototyping of the developed processors. 
First, we modeled the operation and latency of SRAM at the RTL level, with \refTable{tab:latency} reporting the read/write latencies used in the model. 
These results were obtained using the NVSim simulation tool~\cite{dong2012nvsim} at a 45nm process, with different operating voltages of 1.2V for HP-PIM and 0.8V for LP-PIM. 
The 0.8V operating voltage of LP-MRAM, in particular, is based on recent specifications of fabricated STT-MRAM chips~\cite{lee202333, chiu202322nm}. 
We then programmed the developed processors on a Genesys2 Kintex-7 FPGA board~\cite{Genesys2}. 
The operating clock frequency of the prototypes was set to 50 MHz, and memory latencies were scaled according to \refTable{tab:latency}. 
The photograph on the right side of \refFigure{fig:soc} shows the prototype of the HH-PIM processor, while \refTable{tab:fpga_resources} reports the resource consumption of the prototype. 


We verified the correct operation of the FPGA prototypes and measured performance metrics by running benchmark applications, expecting similar results to those observed on an SoC operating at 50 MHz~\cite{han2020developing}. 
For the AI benchmark applications, we used TinyML models based on CNN deep learning backbones such as EfficientNet-B0~\cite{tan2019efficientnet}, MobileNetV2~\cite{sandler2018mobilenetv2}, and ResNet-18~\cite{he2016deep}, which are suitable for edge AI devices. 
We extracted the characteristics and operations of these models, calculated the proportion of PIM operations relative to the total operations, and incorporated these ratios into the benchmark applications. 
The \textit{time slice} for performing the data placement algorithm in each TinyML model was set to allow up to 10 inferences per \textit{time slice}, representing the scenario in which HH-PIM operates at maximum performance. 
\refTable{tab:tinyml_models} reports the characteristics of the models incorporated into the benchmark applications.
Additionally, to reflect dynamic computational load variations encountered at runtime, we configured six workload scenarios as shown in \refFigure{fig:workload-patterns}. 
These scenarios include a consistently low workload pattern (Case 1), a consistently high workload pattern (Case 2), periodic spike patterns (Case 3 and 4), a pulsing pattern with alternating high and low workloads (Case 5), and a random workload pattern (Case 6). 
The spike and pulse patterns simulate realistic scenarios in AI applications on edge devices, where computational demands periodically surge.

\begin{figure*}[t]
\vskip -10pt
\centering
\includegraphics[width=0.98\textwidth]{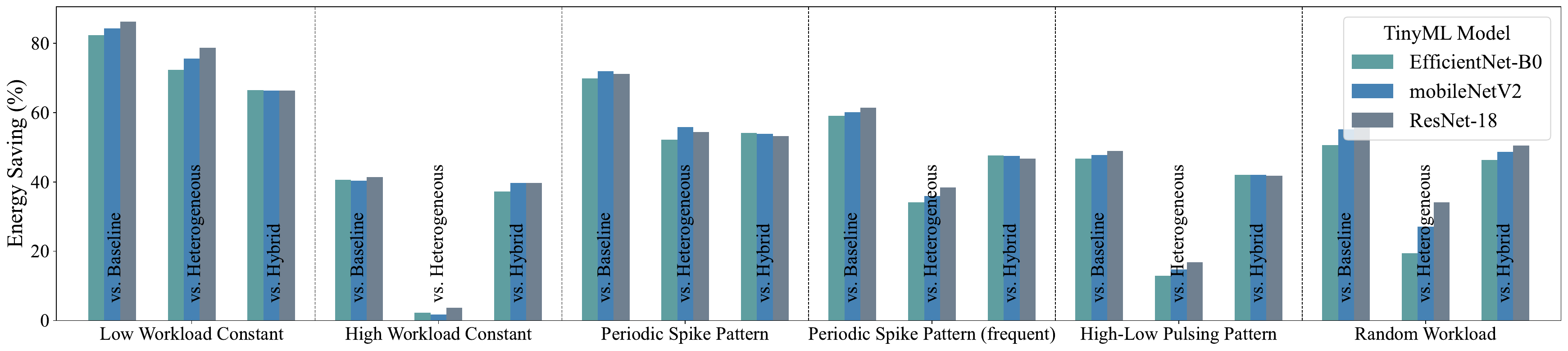}
\vskip -3pt
\caption{Energy savings of the HH-PIM over the Baseline-, Heterogeneous-, and Hybrid-PIM across benchmark scenarios.}
\label{fig:energy_results}
\vskip -6pt
\end{figure*}

To measure the power consumption of the developed processors, we then synthesized them at a 45nm process technology. 
The RTL design, excluding memory, was synthesized using Synopsys Design Compiler with the 45nm Nangate PDK~\cite{NCSU}, while the memory power values were obtained through NVSim simulations at 45nm technology, as previously described. 
\refTable{tab:power} details the power consumption of each memory type. 
These values were subsequently combined with benchmark-specific execution times, as measured from FPGA prototypes, to calculate total energy consumption. 
A detailed analysis of energy consumption, along with energy-saving results across various benchmark scenarios shown in \refFigure{fig:energy_results}, is discussed further in the following section.


\begin{figure}[t]
\vskip 0pt
\centerline{\includegraphics[width=0.48\textwidth]{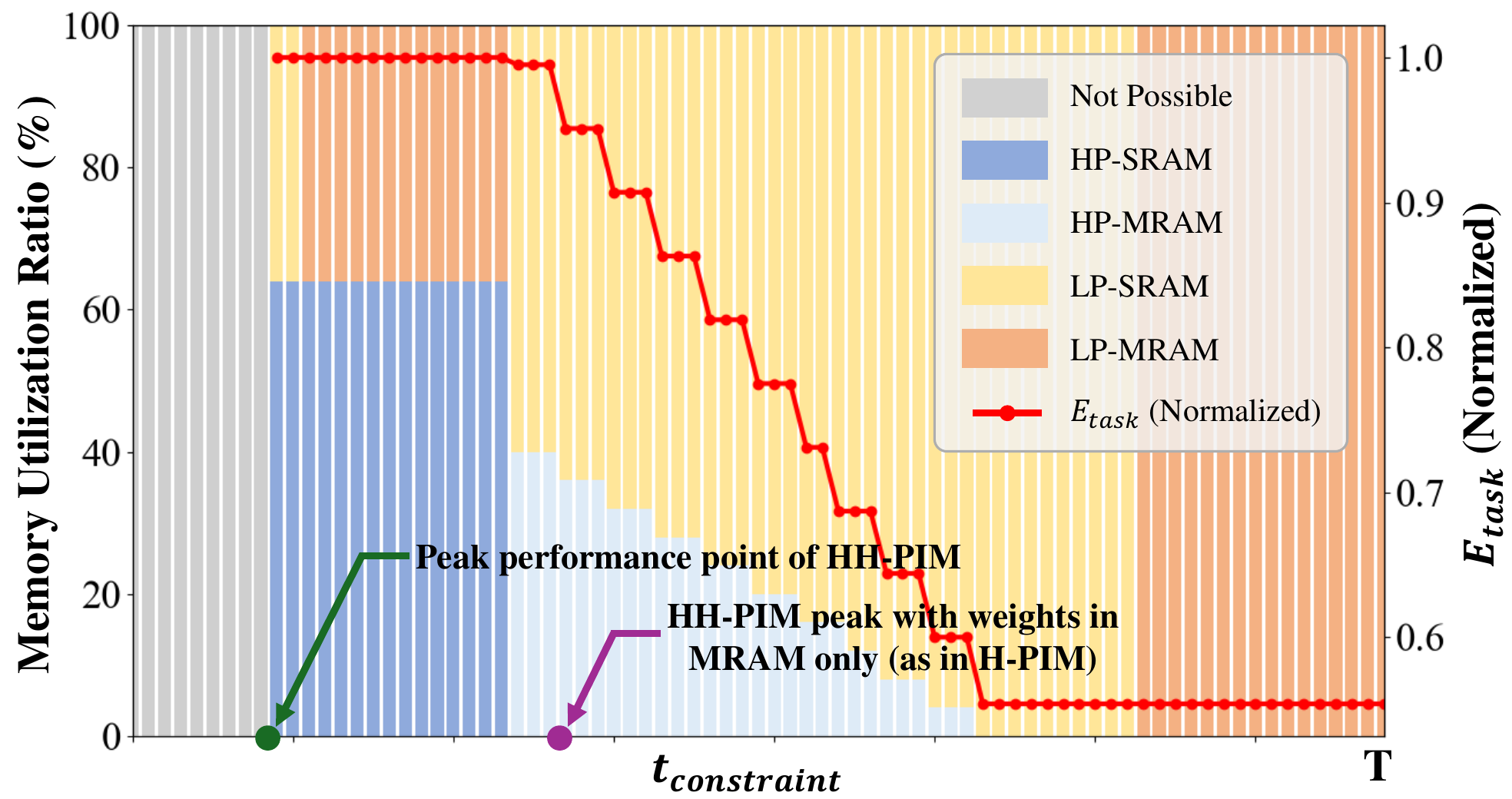}}
\vskip 0pt
\caption{Memory utilization and energy consumption across $t_{constraint}$ within \textit{time slice} ($T$) based on HH-PIM's optimized data placement.}
\label{fig:Data_allo}
\end{figure}

 \subsection{Comprehensive Analysis of Energy Savings}

\refFigure{fig:Data_allo} reports the results of applying the proposed data placement optimization algorithm to benchmark applications executed on HH-PIM. 
The x-axis represents the $t_{constraint}$ within a \textit{time slice} ($T$), while the left y-axis indicates the memory utilization determined by the optimal data placement under the given $t_{constraint}$, and the right y-axis represents the corresponding energy consumption $E_{task}$ of HH-PIM.
The gray region in the graph denotes cases where $t_{constraint}$ is too small, i.e., the required performance is too high for HH-PIM to handle. 
The green dot marks the point where HH-PIM operates at peak performance, satisfying the application's performance requirements. 
At this point, HH-PIM actively utilizes SRAMs for data storage, with the total neural network data stored in a 16:9 ratio between HP-SRAM and LP-SRAM, minimizing PIM module idle times. 
The inference times at this point are measured as 31.06$ms$, 25.71$ms$, and 320.87$ms$ for the EfficientNet-B0, MobileNetV2, and ResNet-18 benchmarks, respectively. 
As expected, energy consumption peaks at this point, and $E_{task}$ is normalized to this value in the graph.
Additionally, the purple dot indicates the peak performance point of HH-PIM when weights are stored only in MRAM (as in previous H-PIMs). 
At this point, the inference times for the EfficientNet-B0, MobileNetV2, and ResNet-18 benchmarks are 44.5$ms$, 36.84$ms$, and 459.74$ms$, respectively. 
This demonstrates that utilizing both SRAM and MRAM for weight storage, as proposed, outperforms the traditional method in terms of performance.


\refFigure{fig:Data_allo} demonstrates the sequential allocation of weight data across memory types as $t_{constraint}$ increases. 
Specifically, the distribution progresses through combinations of HP-SRAM and LP-MRAM, HP-MRAM and LP-SRAM, LP-SRAM alone, and finally LP-MRAM alone. 
This sequence reflects the balance between performance and energy efficiency across memory types. 
In the range where $t_{constraint}$ is longest, all weight data are stored exclusively in LP-MRAM, which operates at minimal power. 
In this configuration, other memory types are deactivated through power-gating, eliminating standby power and maximizing the processor's energy efficiency. 
Consequently, in this highly efficient region, HH-PIM achieves up to a 43.17\% reduction in $E_{task}$ compared to unoptimized data allocation. 
This significant reduction underscores the practical value of the proposed algorithm in real-world edge AI applications. 
Additionally, in the mid-range of $t_{constraint}$, $E_{task}$ exhibits a quasi-linear decline with intermittent plateaus as $t_{constraint}$ increases. 
This pattern reflects the algorithm's capability to shift data progressively to lower-power memory while meeting inference latency constraints. 
These results demonstrate that the proposed solution achieves optimal energy efficiency within the constraints of application inference latency for each $t_{constraint}$, ensuring that HH-PIM can dynamically allocate data to balance energy savings and performance effectively. 



\refFigure{fig:energy_results} presents the energy savings achieved by HH-PIM compared to Baseline-PIM, Hetero.-PIM, and H-PIM (Hybrid-PIM) when executing the benchmark application across 50 \textit{time slices} under various inference pattern scenarios from \refFigure{fig:workload-patterns}.
The best-case scenario for HH-PIM is Case 1 (low workload constant), where the inference pattern remains low across all intervals. 
In this scenario, HH-PIM achieved energy savings of up to 86.23\%, 78.7\%, and 66.5\% compared to Baseline-, Hetero.-, and H-PIM, respectively.
Conversely, the worst-case scenario occurs in Case 2 (high workload constant), where the inference pattern remains consistently high. 
Here, HH-PIM still delivered notable energy savings of up to 41.46\% and 39.69\% compared to Baseline- and H-PIM, respectively. 
However, savings against Hetero.-PIM were limited to 3.72\%, as both HH-PIM and Hetero.-PIM primarily utilize HP-SRAM and LP-SRAM in this scenario, resulting in minimal differences.
Detailed energy savings for the other cases are reported in \refTable{tab:energy_results}.
HH-PIM achieved the highest energy savings over the baseline in ResNet-18, with up to 8.59\% difference across cases.
On average, HH-PIM achieved energy savings of up to 60.43\%, 36.3\%, and 48.58\% compared to Baseline-PIM, Hetero.-PIM, and H-PIM, respectively.


\begin{table}
\vskip 5pt
\centering
\caption{Energy Savings ($ES$) by HH-PIM for Cases 3--6.}
\resizebox{0.92\columnwidth}{!}{
\begin{tabular}{c|ccc}
\toprule
\multirow{2}{*}{\textbf{ \diagbox{}{~~~~~~~~~~$\boldsymbol{ES}$ \textbf{(\%)}\\ ~~over}}} & \textbf{Baseline-} & \textbf{Hetero.-} & \multirow{2}{*}{\textbf{H-PIM}}  \\
& \textbf{PIM} & \textbf{PIM}  \\ \midrule
Case 3: Periodic Spike                    & 72.01    & 55.78    & 54.09 \\ \hline
Case 4: Periodic Spike (frequent)         & 61.46    & 38.38    & 47.60 \\\hline
Case 5: High-Low Pulsing                  & 48.94    & 16.89    & 42.10 \\\hline
Case 6: Random                            & 59.28    & 34.14    & 50.52 \\ \bottomrule
\end{tabular}
}
\label{tab:energy_results}
\vskip 5pt
\end{table}


\section{Conclusion}
In this study, we proposed the HH-PIM architecture designed to efficiently execute AI applications on edge devices. The architecture integrates HP MRAM-SRAM PIM modules and LP MRAM-SRAM PIM modules to achieve a balance between performance and energy efficiency. Additionally, we introduced a data placement optimization algorithm that dynamically allocates data based on computational demand, maximizing energy savings while maintaining application performance requirements.
To validate the effectiveness of the proposed architecture and algorithm, we designed processors incorporating HH-PIM and established comparison baselines with processors equipped with Baseline-, Heterogeneous-, and Hybrid-PIM architectures. These designs were evaluated through FPGA prototyping and power simulations. By running various edge AI application benchmarks, the proposed HH-PIM demonstrated its superiority, achieving up to 60.43\% average energy savings compared to conventional PIM architectures, while meeting application performance requirements.

\bibliographystyle{IEEEtran}
\bibliography{reference}
\end{document}